\documentclass[11pt, a4paper]{article}
\usepackage{amsmath}
\usepackage{amssymb}

\newcommand{\be}{\begin{equation}}
\newcommand{\ee}{\end{equation}}
\newcommand{\bea}{\begin{eqnarray}}
\newcommand{\eea}{\end{eqnarray}}
\newcommand{\ba}{\begin{array}}
\newcommand{\ea}{\end{array}}

\begin{document}
\begin{center}
\baselineskip 20pt
{\Large\bf
Goldstone Modes and Clebsch-Gordan Coefficients
}
\vspace{1cm}

{\large
Xian-Zheng Bai$^a$ \footnote{ E-mail: 15012100098@pku.edu.cn}, Deshan Yang$^b$ \footnote{ E-mail: yangds@ucas.ac.cn}
and
Da-Xin Zhang$^a$\footnote{ E-mail: dxzhang@pku.edu.cn}
}
\vspace{.5cm}

{\baselineskip 20pt \it
a) School of Physics and State Key Laboratory of Nuclear Physics and Technology, \\
Peking University, Beijing 100871, China\\[0.3cm]
b)School of Physical Sciences, University of Chinese Academy of Sciences, Beijing 100049, China}

\vspace{.5cm}

\vspace{1.5cm} {\bf Abstract}
\end{center}
We solve explicitly the Goldstone modes  in spontaneously symmetry breaking
models with  supersymmetry.
We find that, when more than one fields or representations contribute to the symmetry breaking,
there exist
identities among the Clebsch-Gordan coefficients which can be used as
consistent checks on the calculations.

\section{Introduction}

Spontaneously Symmetry Breaking (SSB)\cite{ssb1,ssb2,ssb3} is a very important concept in setting up
the Standard Model (SM)\cite{sm1,sm2,sm3}.
It is also needed in building  models of the Grand Unified Theories (GUTs)\cite{gut1,gut2} and the
Supersymmetric (SUSY) GUTs \cite{sgut1,so10a,so10b}.
When SSB happens, the neutral components of scalar fields develop Vacuum Expectation Values (VEVs)
to break the symmetry, while the other components corresponding to
the broken generators of the symmetry become massless Goldstone bosons.
In gauge theory, these Goldstones become the longitudinal components of the gauge fields to make
them massive\cite{higgs1,higgs2,higgs3,higgs4,higgs5}, and we will call these scalar fields the Higgs fields.
Furthermore,
when the theory is SUSY,  the Goldstinos, which are the fermionic partners of the Goldstones,
are also massless and, in gauge theory, give masses to the gauginos, the fermionic partners of the
gauge fields.

In GUT models, especially in the SUSY GUT models beyond the minimal SU(5),
there are generally several Higgs fields which break a same
symmetry. The SO(10) examples can be found in  \cite{he,lee,sato,np597,so10c,so10e,so10d,so10f,fuku,np711,np757,so10g,np857,np882,nath2015,czy}.
However, the SSB in these models has not been given a general treatment.
The contents of the Goldstones have not been studied systematically.
In the literature, to check whether the SSB happens,
numerically studies are common \cite{sato,so10e,fuku,so10g}. In this paper,
we will show the contents of the Goldstones in SUSY models so that numerical tests are not needed.

The study of SSB can be easier in the supersymmetric models, in which
the Goldstones and the Goldstinos share the same contents in  the case that several Higgs fields contribute to the same SSB.
The  Goldstinos satisfy the zero eigenvalue equations which are linear in the mass matrices. Such equations are
easier to solve than those for the Goldstones which are quadratic in the mass mactirces.

In this work, we will study the Goldstone modes in the general SUSY models.
We will show explicitly that for a Goldstone mode for the SSB,
besides the constituent in any representation is proportional to the relevant VEV
in the same representation
and does not depend directly on the parameters of the models,
only the Clebsch-Gordan coefficients (CGCs) are relevant.
We find some non-trivial identities among the CGCs.
We will first show the method in the cases that only real representations are involved,
then generalize to the  cases with complex representations,
and present some identities in the simplest cases of SU(2) whose CGCs were
well studied in the literature. Although most of identities for SU(2) CGCs in the literature are related to the angular momentum theory in Quantum Mechanics, a series of identities among CGCs of SU(2) group that we present in this paper are new, due to the fact that they are obtained through the new perspective to the study of SSB.

\section{SSB for All Real Representations}
In studying the Goldstone modes we need to identify their related  broken symmetries.
For example, in the SO(10) models,
the Goldstones with the SM representations $(1,1,1) + c.c.$ are relevant to the SSB of
the SO(10) subgroup $SU(2)_R$ into $U(1)_{I_{3R}}$,
$(3,1,\frac{2}{3}) + c.c.$ to $SU(4)_C$ into $SU(3)_C\otimes U(1)_{B-L}$,
$(3,2, -\frac{5}{6})+{ c.c.}$ to $SU(5)$ into the SM, and $(3,2,\frac{1}{6})+{ c.c.}$
to the flipped $SU(5)$ into the SM.
We will  study the SSB of these subgroups instead of the SO(10),
since in a complete representation of SO(10),
there might be more than one SM singlets which can develop VEVs to break different SO(10) subgroups,
and there might be more than one Goldstone components which have the same SM representations
but cannot be distinguished if the broken subgroups are not identified.

Here we consider the SSB of the $G_1 \rightarrow G_2$ within SUSY models,
so that within a representation of $G_1$ all the components are distinguished by their representations under $G_2$.
To start with, we consider the simplest case where all representations are real.
The general Higgs superpotential can be written as
\begin{equation}
W = \frac{1}{2} \sum_{I,i} m_I (I_i)^2
+ \sum_{IJK,ijk} {\lambda}^{IJK} I_i J_j K_k,\label{real1}
\end{equation}
where $I,J,K$ denote the superfields, $i,j,k$ denote the representations under $G_1$.
The superscripts in $\lambda^{IJK}$ is used to ensure all possible terms
are included when there exist more than one fields in the same representations.

Under $G_2$, the couplings in (\ref{real1}) are of the forms
\begin{equation}
{\lambda}^{IJK} I_{i_a} J_{j_b} K_{k_c} C^{ijk}_{abc}, \label{real2}
\end{equation}
where $a,b,c$ are the representations under $G_2$, and $C^{ijk}_{abc}$ is the Clebsch-Gordan coefficient (CGC)
in the sense that the coupling among the field components $I_{i_a}, J_{j_b}, K_{k_c}$ is definite.
In the SSB a singlet of $G_2$
can have a VEV  which will be denoted as, {\it e.g.}, ${I}_{i_0}$ where $i_0$ stands for a $G_2$ singlet.
SUSY requires the F-flatness condition for the singlet,
\begin{equation}
0\equiv F_{I_{i_0}} = \left\langle \frac{\partial W}{\partial I_{i_0}} \right\rangle
= m_I  I_{i_0} + \sum_{JK,jk} \lambda^{IJK}  J_{j_0}  K_{k_0} C^{ijk}_{000}.
\label{realF}
\end{equation}

The Goldstinos
in the SSB will be denoted by representations $\alpha,\overline{\alpha}$ under $G_2$.
The mass matrix element for the Goldstinos is
\begin{equation}
M^{IL}_{{i_\alpha}{l_{\bar{\alpha}}}}
= \left\langle \frac{\partial^2 W}{\partial I_{i_\alpha} \partial L_{l_{\bar{\alpha}}}} \right\rangle
= m_I \delta_{IL} \delta_{il}
+ \sum_{JK,jk} \lambda^{IJK}  (K_{k_0} C^{ijk}_{{\alpha}{{\bar{\alpha}}}{0}}\delta_{JL}\delta_{jl}
+J_{j_0} C^{ijk}_{{\alpha}{0}{{\bar{\alpha}}}}\delta_{KL}\delta_{kl}).
\end{equation}
$\widehat{G}^I_{i_{\bar{\alpha}}}$ denotes the component of the Goldstino corresponding to
the $\bar{\alpha}$ under $G_2$, and it satisfies
the zero-eigenvalue equation
\begin{equation}
0\equiv \sum_{L,l} M^{IL}_{{i_\alpha}{l_{\bar{\alpha}}}} \widehat{G}^L_{l_{\bar{\alpha}}}
= m_I \widehat{G}^I_{i_{\bar{\alpha}}} + \sum_{JK,jk} \lambda^{IJK}
(\widehat{G}^J_{j_{\bar{\alpha}}} K_{k_0} C^{ijk}_{{\alpha}{{\bar{\alpha}}}{0}}
+\widehat{G}^K_{k_{\bar{\alpha}}} J_{j_0} C^{ijk}_{{\alpha}{0}{{\bar{\alpha}}}}).  \label{realM}
\end{equation}
Eliminating $m_I$ in (\ref{realF},\ref{realM}), it gives
\begin{equation}
0 = \widehat{G}^I_{i_{\bar{\alpha}}} \sum\limits_{JK,jk} \lambda^{IJK} J_{j_0} K_{k_0} C^{ijk}_{000}
- I_{i_0} \sum\limits_{JK,jk} \lambda^{IJK}
(\widehat{G}^J_{j_{\bar{\alpha}}}K_{k_0} C^{ijk}_{{\alpha}{{\bar{\alpha}}}{0}}
+\widehat{G}^K_{k_{\bar{\alpha}}}J_{j_0} C^{ijk}_{{\alpha}{0}{{\bar{\alpha}}}}).
\end{equation}
It is now clear that if a representation of $G_1$ does not contain a $G_2$ singlet,
even if it may contain the same representations as the Goldstinos,
it does not contribute to the Goldstinos since the $m_I$ term cannot be eliminated unless
multiplied by zero.

The superpotential parameters can be arbitrary so that we can focus on a specified coupling
$\lambda^{IJK}$, and the summations are unnecessary.
Furthermore, denoting
\begin{equation}
T^{Ii}_{\bar{\alpha}} = \frac{\widehat{G}^I_{i_{\bar{\alpha}}}}{I_{i_0}}, ~\cdots,\label{T1}
\end{equation}
we have
\begin{equation}
0 = T^{Ii}_{\bar{\alpha}} \lambda^{IJK} J_{j_0}  K_{k_0} C^{ijk}_{000}
- T^{Jj}_{\bar{\alpha}} \lambda^{IJK}  J_{j_0} K_{k_0} C^{ijk}_{{\alpha}{{\bar{\alpha}}}{0}}
- T^{Kk}_{\bar{\alpha}} \lambda^{IJK} J_{j_0} K_{k_0} C^{ijk}_{{\alpha}{0}{{\bar{\alpha}}}},\nonumber
\end{equation}
therefore the nonzero factor $\lambda^{IJK} J_{j_0} K_{k_0}$ can be eliminated.
Reiterating the same operations for $J,K$ gives
\begin{equation}
0 = \left( \begin{array}{ccc}
- C^{ijk}_{000} &  C^{ijk}_{{\alpha}{{\bar{\alpha}}}{0}} &  C^{ijk}_{{\alpha}0{{\bar{\alpha}}}} \\ \\
C^{ijk}_{{{\bar{\alpha}}}{\alpha}{0}} & - C^{ijk}_{000} &
C^{ijk}_{0{\alpha}{{\bar{\alpha}}}} \\ \\
C^{ijk}_{{{\bar{\alpha}}} 0 {\alpha}} &  C^{ijk}_{0{{\bar{\alpha}}}{\alpha}} &
- C^{ijk}_{000}
\end{array}
\right)
\left(
\begin{array}{c}
T^{Ii}_{\bar{\alpha}} \\ \\ T^{Jj}_{\bar{\alpha}} \\ \\ T^{Kk}_{\bar{\alpha}}
\end{array}
\right), \label{19a}
\end{equation}
and a similar result can be given for the Goldstino in $\alpha$.
Taking all different couplings $\lambda^{IJK}$ will setup all relations among the Goldstino components.
This determines the Goldstino contents up to an overall normalization.

There are several interesting implications following (\ref{19a}).
The Goldstino, and thus the Goldstone, have components proportional to the VEVs in the $G_1$ representations
(see (\ref{T1}))
and are independent of any masses or couplings.
In general gauge theories with SSB, the same findings follow the eliminations
of the kinematic mixing between the gauge fields and the derivatives
of the Goldstones. The CGC dependence are what we get in the present approach
in the SUSY models.
Also, it implies the identity
\begin{equation}
 \left| \begin{array}{ccc}
- C^{ijk}_{000} &  C^{ijk}_{{\alpha}{{\bar{\alpha}}}{0}} &  C^{ijk}_{{\alpha}0{{\bar{\alpha}}}} \\ \\
C^{ijk}_{{{\bar{\alpha}}}{\alpha}{0}} & - C^{ijk}_{000} &
C^{ijk}_{0{\alpha}{{\bar{\alpha}}}} \\ \\
C^{ijk}_{{{\bar{\alpha}}} 0 {\alpha}} &  C^{ijk}_{0{{\bar{\alpha}}}{\alpha}} &
- C^{ijk}_{000}
\end{array}
\right|=0,
\end{equation}
for the representations  which may serve as Goldstinos.
There are also relations among the CGCs when different couplings have a same superfield in common.
In general gauge theories,
the constituents of the representations in the Goldstones
are also proportional to the matrix elements of
the broken generators. For high  representations in complicated groups,
the explicit constructions of these generators are not available and thus
independent checks of the Goldstone modes are lack.
In the present approach,
(8) can be used as independent checks of the calculated CGCs and the Goldstones.

Special cases are also needed to be discussed.

\noindent
1) For $J = K$ and $j = k$, we have
\begin{equation}
0 = \left( \begin{array}{cc}
- C^{ijj}_{000} &  ~~2C^{ijj}_{{\alpha}{{\bar{\alpha}}}{0}}  \\ \\
C^{ijj}_{{{\bar{\alpha}}}{\alpha}{0}} & ~~C^{ijj}_{0{\alpha}{{\bar{\alpha}}}}
-  C^{ijj}_{000}
\end{array}
\right)
\left(
\begin{array}{c}
T^{Ii}_{\bar{\alpha}} \\ \\ T^{Ij}_{\bar{\alpha}}
\end{array}
\right).
\end{equation}

\noindent
2) For $I = J = K$ and $i = j = k$, the result is
\begin{equation}
C^{iii}_{000} =2 C^{iii}_{{\alpha}0{{\bar{\alpha}}}},
\end{equation}
under the condition $C^{iii}_{000}\neq 0$.

\noindent
3) If in a coupling $\lambda^{IJK}_{ijk}$, $k$ doesn't contain $\alpha,\overline{\alpha}$,
then
\begin{equation}
0 = \left( \begin{array}{cc}
- C^{ijk}_{000} &  C^{ijk}_{{\alpha}{{\bar{\alpha}}}{0}} \\ \\
C^{ijk}_{{{\bar{\alpha}}}{\alpha}{0}} & - C^{ijk}_{000}
\end{array}
\right)
\left(
\begin{array}{c}
T^{Ii}_{\bar{\alpha}} \\ \\ T^{Jj}_{\bar{\alpha}}
\end{array}
\right).
\end{equation}

\section{SSB with Complex Representations}

Supposing that all the  representations are complex, the general superpotential can be written as
\begin{equation}
W =  \sum_{I,i} m_I I_i \bar{I}_{\bar i}
+ \sum_{IJK,ijk}{\lambda}^{IJK} I_i J_j K_k
+ \sum_{\bar{I}\bar{J}\bar{K},\bar{i}\bar{j}\bar{k}}
{\lambda}^{\bar{I}\bar{J}\bar{K}}
\bar{I}_{\bar{i}}\bar{J}_{\bar{j}}\bar{K}_{\bar{k}}\,.\label{complex1}
\end{equation}
Similar to the case of real representations, all the complex representations need to contain
singlets of $G_2$. Otherwise, they will not contribute to the Goldstone modes.
However, $i$ (or $\bar{i},\cdots$) may contain
either Goldstino  $\alpha$, or $\overline\alpha$, or both.
Here we assume that all representations contain both $\alpha$ and $\overline{\alpha}$ at first.

Since the arbitrariness of the parameters,  it is sufficient to
consider one pair of ${\lambda}^{IJK}$ and ${\lambda}^{\bar{I}\bar{J}\bar{K}}$.
Denoting
\begin{equation}
T^{Ii}_{\bar{\alpha}} = \frac{\widehat{G}^I_{i_{\bar{\alpha}}}}{{I}_{i_0}}, ~~~
T^{\bar{I}\bar{i}}_{\bar{\alpha}} = \frac{\widehat{G}^{\bar{I}}_{\bar{i}_{\bar{\alpha}}}}{{\bar{I}}_{\bar{i}_0}},
~~\cdots.\label{cT}
\end{equation}
The mass matrix is
\\ \\
\begin{minipage}{16cm}
	{\bf c:}
	$\bar{I}_{\bar{i}_{\bar{\alpha}}}$,
	$\bar{J}_{\bar{j}_{\bar{\alpha}}}$,
	$\bar{K}_{\bar{k}_{\bar{\alpha}}}$,
	$I_{i_{\bar{\alpha}}}$,
	$J_{j_{\bar{\alpha}}}$,
	$K_{k_{\bar{\alpha}}}$\\[.15cm]
	{\bf r:}
	$I_{i_\alpha}$,
	$J_{j_\alpha}$,
	$K_{k_\alpha}$,
	$\bar{I}_{\bar{i}_\alpha}$,
	$\bar{J}_{\bar{j}_\alpha}$,
	$\bar{K}_{\bar{k}_\alpha}$\\[.2cm]
	\begin{equation}
	\left(
	\begin{array}{cccccc}
		\begin{array}{c}
			m_I \\ \\
			0 \\ \\
			0 \\ \\
			0 \\ \\
			\lambda^{\bar{I}\bar{J}\bar{K}} \bar{K}_{\bar{k}_0} C^{\bar{i}\bar{j}\bar{k}}_{\bar{\alpha} \alpha 0} \\ \\
			\lambda^{\bar{I}\bar{J}\bar{K}} \bar{J}_{\bar{j}_0} C^{\bar{i}\bar{j}\bar{k}}_{\bar{\alpha} 0 \alpha}
		\end{array}
		\begin{array}{c}
			0 \\ \\
			m_J \\ \\
			0 \\ \\
			\lambda^{\bar{I}\bar{J}\bar{K}} \bar{K}_{\bar{k}_0} C^{\bar{i}\bar{j}\bar{k}}_{\alpha \bar{\alpha} 0} \\ \\
			0 \\ \\
			\lambda^{\bar{I}\bar{J}\bar{K}} \bar{I}_{\bar{i}_0} C^{\bar{i}\bar{j}\bar{k}}_{0 \bar{\alpha} \alpha}
		\end{array}
		\begin{array}{c}
	0 \\ \\
	0 \\ \\
		m_K \\ \\
			\lambda^{\bar{I}\bar{J}\bar{K}} \bar{J}_{\bar{j}_0} C^{\bar{i}\bar{j}\bar{k}}_{\alpha 0 \bar{\alpha}} \\ \\
			\lambda^{\bar{I}\bar{J}\bar{K}} \bar{I}_{\bar{i}_0} C^{\bar{i}\bar{j}\bar{k}}_{0 \alpha \bar{\alpha}} \\ \\
			0
		\end{array}
		\begin{array}{c}
			0 \\ \\
			\lambda^{IJK} K_{k_0} C^{ijk}_{\bar{\alpha} \alpha 0} \\ \\
			\lambda^{IJK} J_{j_0} C^{ijk}_{\bar{\alpha} 0 \alpha} \\ \\
			m_I \\ \\
			0 \\ \\
			0
		\end{array}
		\begin{array}{c}
			\lambda^{IJK} K_{k_0} C^{ijk}_{\alpha \bar{\alpha} 0} \\ \\
			0 \\ \\
			\lambda^{IJK} I_{i_0} C^{ijk}_{0 \bar{\alpha} \alpha}  \\ \\
			0 \\ \\
			m_J \\ \\
			0
		\end{array}
		\begin{array}{c}
			\lambda^{IJK} J_{j_0} C^{ijk}_{\alpha 0 \bar{\alpha} } \\ \\
			\lambda^{IJK} I_{i_0} C^{ijk}_{0 \alpha \bar{\alpha}} \\ \\
			0 \\ \\
			0 \\ \\
			0 \\ \\
			m_K
		\end{array}
	\end{array}
	\right).
	\end{equation}
\end{minipage}\\[.2cm]
Acting the mass matrix on the Goldstino $({\bar{I}}_{\bar{i}_0}T^{\bar{I}}_{\bar{i}_{\bar{\alpha}}}, \cdots)^{\rm T}$,
and eliminating the $m_{I,J,K}$s through the
F-flatness conditions which are
\be
	0\equiv F^I_{i_0} = m_I \bar{I}_{\bar{i}_0} + \lambda^{IJK}  J_{j_0} K_{k_0}
	C^{ijk}_{000}, ~~~~
	0\equiv  F^{\bar{I}}_{\bar{i}_0} = m_I I_{i_0} +\lambda^{\bar{I}\bar{J}\bar{K}} \bar{J}_{\bar{j}_0} \bar{K}_{\bar{k}_0} C^{\bar{i}\bar{j}\bar{k}}_{000}, \cdots ,
\ee	
we obtain
\\
	\begin{equation}
	0 =\left(
	\begin{array}{cccccc}
	\begin{array}{c}
	- C^{ijk}_{000} \\ \\
	0 \\ \\
	0 \\ \\
	0 \\ \\
	 C^{\bar{i}\bar{j}\bar{k}}_{\bar{\alpha} \alpha 0} \\ \\
	 C^{\bar{i}\bar{j}\bar{k}}_{\bar{\alpha} 0 \alpha}
	\end{array}
	\begin{array}{c}
	0 \\ \\
	- C^{ijk}_{000} \\ \\
	0 \\ \\
	 C^{\bar{i}\bar{j}\bar{k}}_{\alpha \bar{\alpha} 0} \\ \\
	0 \\ \\
	 C^{\bar{i}\bar{j}\bar{k}}_{0 \bar{\alpha} \alpha}
	\end{array}
	\begin{array}{c}
	0 \\ \\
	0 \\ \\
	- C^{ijk}_{000} \\ \\
	 C^{\bar{i}\bar{j}\bar{k}}_{\alpha 0 \bar{\alpha}} \\ \\
	 C^{\bar{i}\bar{j}\bar{k}}_{0 \alpha \bar{\alpha}} \\ \\
	0
	\end{array}
	\begin{array}{c}
	0 \\ \\
	 C^{ijk}_{\bar{\alpha} \alpha 0} \\ \\
	 C^{ijk}_{\bar{\alpha} 0 \alpha} \\ \\
	- C^{\bar{i}\bar{j}\bar{k}}_{000} \\ \\
	0 \\ \\
	0
	\end{array}
	\begin{array}{c}
	 C^{ijk}_{\alpha \bar{\alpha} 0} \\ \\
	0 \\ \\
	 C^{ijk}_{0 \bar{\alpha} \alpha}  \\ \\
	0 \\ \\
	- C^{\bar{i}\bar{j}\bar{k}}_{000} \\ \\
	0
	\end{array}
	\begin{array}{c}
	 C^{ijk}_{\alpha 0 \bar{\alpha} } \\ \\
	 C^{ijk}_{0 \alpha \bar{\alpha}} \\ \\
	0 \\ \\
	0 \\ \\
	0 \\ \\
	- C^{\bar{i}\bar{j}\bar{k}}_{000}
	\end{array}
	\end{array}
	\right)
	\left(
	\begin{array}{c}
	T^{\bar{I}}_{\bar{i}_{\bar{\alpha}}} \\ \\
	T^{\bar{J}}_{\bar{j}_{\bar{\alpha}}} \\ \\
	T^{\bar{K}}_{\bar{k}_{\bar{\alpha}}} \\ \\
	T^I_{i_{\bar{\alpha}}} \\ \\
	T^J_{j_{\bar{\alpha}}} \\ \\
	T^K_{k_{\bar{\alpha}}}
	\end{array}
	\right).\label{compeqn}
	\end{equation}
One can see now that only the CGCs, together with the VEVs, are relevant in the Goldstone modes.

There are special cases to be discussed below.

\noindent
1) When $K$ is real, the superpotential is	
\begin{equation}
W = m_I I_i \bar{I}_{\bar{i}} + m_J J_j \bar{J}_{\bar{j}} + \frac{1}{2} m_K K_k^2
+ \lambda^{IJK} I_i J_j K_k + \lambda^{\bar{I}\bar{J}K} \bar{I}_{\bar{i}} \bar{J}_{\bar{j}} K_k,
\end{equation}
the F-flatness conditions for $I_{i_0},J_{j_0}, {\bar{I}}_{\bar{i}_0}, {\bar{J}}_{\bar{j}_0}$
give a relation
\begin{equation}
\frac{\lambda^{\bar{I}\bar{J}K} {\bar{I}}_{\bar{i}_0}{\bar{J}}_{\bar{j}_0}}{\lambda^{IJK}I_{i_0}J_{j_0}}
=\frac{C^{ijk}_{000}}{C^{\bar{i}\bar{j}k}_{000}}\equiv X.
\end{equation}
The mass matrix is
\\ \\
\begin{minipage}{16cm}
	{\bf c:}
	$\bar{I}_{\bar{i}_{\bar{\alpha}}}$,
	$\bar{J}_{\bar{j}_{\bar{\alpha}}}$,
	$K_{k_{\bar{\alpha}}}$,
	$I_{i_{\bar{\alpha}}}$,
	$J_{j_{\bar{\alpha}}}$\\[.15cm]
	{\bf r:}
	$I_{i_\alpha}$,
	$J_{j_\alpha}$,
	$K_{k_\alpha}$,
	$\bar{I}_{\bar{i}_\alpha}$,
	$\bar{J}_{\bar{j}_\alpha}$\\[.2cm]
	\begin{equation}
	\left(
	\begin{array}{ccccc}
	\begin{array}{c}
	m_I \\ \\
	0 \\ \\
	\lambda^{\bar{I}\bar{J}K} \bar{J}_{\bar{j}_0} C^{\bar{i}\bar{j}k}_{\bar{\alpha} 0 \alpha} \\ \\
	0 \\ \\
	\lambda^{\bar{I}\bar{J}K} K_{k_0} C^{\bar{i}\bar{j}k}_{\bar{\alpha} \alpha 0}
	\end{array}
	\begin{array}{c}
	0 \\ \\
	m_J \\ \\
	\lambda^{\bar{I}\bar{J}K} \bar{I}_{\bar{i}_0} C^{\bar{i}\bar{j}k}_{0 \bar{\alpha} \alpha} \\ \\
	\lambda^{\bar{I}\bar{J}K} K_{k_0} C^{\bar{i}\bar{j}k}_{\alpha \bar{\alpha} 0} \\ \\
	0
	\end{array}
	\begin{array}{c}
	\lambda^{IJK} J_{j_0} C^{ijk}_{\alpha 0 \bar{\alpha}} \\ \\
	\lambda^{IJK} I_{i_0} C^{ijk}_{0 \alpha \bar{\alpha}} \\ \\
	m_K \\ \\
	\lambda^{\bar{I}\bar{J}K} \bar{J}_{\bar{j}_0} C^{\bar{i}\bar{j}k}_{\alpha 0 \bar{\alpha}} \\ \\
	\lambda^{\bar{I}\bar{J}K} \bar{I}_{\bar{i}_0} C^{\bar{i}\bar{j}k}_{0 \alpha \bar{\alpha} }
	\end{array}
	\begin{array}{c}
	0 \\ \\
	\lambda^{IJK} K_{k_0} C^{ijk}_{\bar{\alpha} \alpha 0} \\ \\
	\lambda^{IJK} J_{j_0} C^{ijk}_{\bar{\alpha} 0 \alpha} \\ \\
	m_I \\ \\
	0
	\end{array}
	\begin{array}{c}
	\lambda^{IJK} K_{k_0} C^{ijk}_{\alpha \bar{\alpha} 0} \\ \\
	0 \\ \\
	\lambda^{IJK} I_{i_0} C^{ijk}_{0 \bar{\alpha} \alpha}  \\ \\
	0 \\ \\
	m_J
	\end{array}
	\end{array}
	\right).
	\end{equation}
\end{minipage}\\[.2cm]	
Eliminating the parameters $m_{I,J}$ gives
\\ \\
\begin{minipage}{16cm}
	\begin{equation}
	0 =\left(
	\begin{array}{ccccc}
	\begin{array}{c}
	- C^{ijk}_{000} \\ \\
	0 \\ \\
	X C^{\bar{i}\bar{j}{k}}_{\bar{\alpha} 0 \alpha} \\ \\
	0 \\ \\
	 C^{\bar{i}\bar{j}{k}}_{\bar{\alpha} \alpha 0}
	\end{array}
	\begin{array}{c}
	0 \\ \\
	- C^{ijk}_{000} \\ \\
	X C^{\bar{i}\bar{j}{k}}_{0 \bar{\alpha} \alpha} \\ \\
	 C^{\bar{i}\bar{j}{k}}_{\alpha \bar{\alpha} 0} \\ \\
	0
	\end{array}
	\begin{array}{c}
	C^{ijk}_{\alpha 0 \bar{\alpha} } \\ \\
	 C^{ijk}_{0 \alpha \bar{\alpha}} \\ \\
	-2 C^{ijk}_{000} \\ \\
	 C^{\bar{i}\bar{j}{k}}_{\alpha 0 \bar{\alpha}} \\ \\
	 C^{\bar{i}\bar{j}{k}}_{0 \alpha \bar{\alpha}}
	\end{array}
	\begin{array}{c}
	0 \\ \\
	 C^{ijk}_{\bar{\alpha} \alpha 0} \\ \\
	 C^{ijk}_{\bar{\alpha} 0 \alpha} \\ \\
	- C^{\bar{i}\bar{j}{k}}_{000} \\ \\
	0
	\end{array}
	\begin{array}{c}
	 C^{ijk}_{\alpha \bar{\alpha} 0} \\ \\
	0 \\ \\
	 C^{ijk}_{0 \bar{\alpha} \alpha}  \\ \\
	0 \\ \\
	- C^{\bar{i}\bar{j}{k}}_{000}
	\end{array}	
\end{array}
\right)
	\left(
	\begin{array}{c}
	T^{\bar{I}}_{\bar{i}_{\bar{\alpha}}} \\ \\
	T^{\bar{J}}_{\bar{j}_{\bar{\alpha}}} \\ \\
	T^{{K}}_{{k}_{\bar{\alpha}}} \\ \\
	T^I_{i_{\bar{\alpha}}} \\ \\
	T^J_{j_{\bar{\alpha}}}
	\end{array}
	\right).
	\end{equation}
\end{minipage}\\[.2cm]		
	\\ \\

\noindent
2) If $I = J$ when $K$ is real, the superpotential is	
\begin{equation}
W = m_I I_i \bar{I}_{\bar{i}} + \frac{1}{2} m_K K_k^2
+ \lambda^{IIK} I_i I_i K_k + \lambda^{\bar{I}\bar{I}K} \bar{I}_{\bar{i}} \bar{I}_{\bar{i}} K_k.
\end{equation}	
The mass matrix is
\\ \\
\begin{minipage}{16cm}
	{\bf c:}
	$\bar{I}_{\bar{i}_{\bar{\alpha}}}$,
	$K_{k_{\bar{\alpha}}}$,
	$I_{i_{\bar{\alpha}}}$\\[.15cm]
	{\bf r:}
	$I_{i_\alpha}$,
	$K_{k_\alpha}$,
	$\bar{I}_{\bar{i}_\alpha}$\\[.2cm]
	\begin{equation}
	\left(
	\begin{array}{ccc}
	\begin{array}{c}
	m_I \\ \\
	\lambda^{\bar{I}\bar{I}K} \bar{I}_{\bar{i}_0} C^{\bar{i}\bar{i}k}_{\bar{\alpha} 0 \alpha} \\ \\
	\lambda^{\bar{I}\bar{I}K} K_{k_0} C^{\bar{i}\bar{i}k}_{\alpha \bar{\alpha} 0} \\ \\
	\end{array}
	\begin{array}{c}
	\lambda^{IIK} I_{i_0} C^{iik}_{\alpha 0 \bar{\alpha}}  \\ \\
	m_K \\ \\
	\lambda^{\bar{I}\bar{I}K} \bar{I}_{\bar{i}_0} C^{\bar{i}\bar{i}k}_{\alpha 0 \bar{\alpha}}
	\end{array}
	\begin{array}{c}
	\lambda^{IIK} K_{k_0} C^{iik}_{\alpha \bar{\alpha} 0} \\ \\
	\lambda^{IIK} I_{i_0} C^{iik}_{\bar{\alpha} 0 \alpha}  \\ \\
	m_I
	\end{array}
	\end{array}
	\right),\nonumber
	\end{equation}
\end{minipage}\\[.2cm]			
and the F-flatness conditions are
\begin{eqnarray}
F^I_{i_0} &=& m_I \bar{I}_{\bar{i}_0} + \lambda^{IIK} I_{i_0} K_{k_0}
C^{iik}_{000}, \nonumber\\
F^{\bar{I}}_{\bar{i}_0} &=& m_I I_{i_0} +\lambda^{\bar{I}\bar{I}K} \bar{I}_{\bar{i}_0} K_{k_0} C^{\bar{i}\bar{i}k}_{000}, \nonumber\\
F^K_{k_0} &=& m_K K_{k_0} + \lambda^{IIK} I_{i_0} I_{i_0} C^{iik}_{000}
+ \lambda^{\bar{I}\bar{I}K} \bar{I}_{\bar{i}_0} \bar{I}_{\bar{i}_0} C^{\bar{i}\bar{i}k}_{000}.
\end{eqnarray}	
so
\\
\begin{minipage}{16cm}
\be
0=\left(
\ba{ccc}
\ba{c}
-C^{iik}_{000} \\ \\
XC^{\bar{i}\bar{i}k}_{\bar{\alpha}0\alpha}\\ \\
C^{\bar{i}\bar{i}k}_{\bar{\alpha}\alpha0}
\ea
\ba{c}
C^{{i}{i}k}_{\alpha 0\bar{\alpha}}\\ \\
-C^{iik}_{000} \\ \\
C^{\bar{i}\bar{i}k}_{\bar{\alpha}0\alpha}
\ea
\ba{c}
C^{{i}{i}k}_{\alpha\bar{\alpha} 0}\\ \\
C^{{i}{i}k}_{\bar{\alpha}0\alpha}\\ \\
-C^{\bar{i}\bar{i}k}_{000}
\ea
\ea
\right)
\left(
\ba{c}
T^{\bar{I}i}_{\bar{\alpha}} \\ \\ T^{Kk}_{\bar{\alpha}}\\ \\T^{Ii}_{\bar{\alpha}}
\ea
\right),
\ee
\end{minipage}\\[.2cm]		
\\ \\
where $X=\frac{C^{\bar{i}\bar{i}k}_{000}}{C^{iik}_{000}}$.

\noindent
3) If $I = \bar{J}$ when $K$ is real, the superpotential is
\begin{equation}
W = m_I I_i \bar{I}_{\bar{i}}  + \frac{1}{2} m_K K_k^2
+ \lambda^{I\bar{I}K} I_i \bar{I}_j K_k,
\end{equation}
we get\\
\be
0=\left(
\ba{ccc}
\ba{c}
-C^{i\bar{i}k}_{000} \\ \\
C^{i\bar{i}k}_{0\bar{\alpha}\alpha}\\ \\
0
\ea
\ba{c}
C^{i\bar{i}k}_{\alpha 0\bar{\alpha}}\\ \\
-C^{i\bar{i}k}_{000} \\ \\
C^{i\bar{i}k}_{0\alpha\bar{\alpha}}
\ea
\ba{c}
0\\ \\
C^{i\bar{i}k}_{\bar{\alpha}0\alpha}\\ \\
-C^{i\bar{i}k}_{000}
\ea
\ea
\right)
\left(
\ba{c}
T^{\bar{I}i}_{\bar{\alpha}} \\ \\ T^{Kk}_{\bar{\alpha}}\\ \\T^{Ii}_{\bar{\alpha}}
\ea
\right),
\ee

\noindent
4) If $I=J=K$, the superpotential is
\begin{equation}
W = m_I I_i \bar{I}_{\bar{i}}
+ \lambda^{III} I_i^3 + \lambda^{\bar{I}\bar{I}\bar{I}} \bar{I}_{\bar{i}}^3,
\end{equation}	
we get
\begin{minipage}{16cm}
\be
0=\left(
\ba{ccc}
\ba{c}
-C^{iii}_{000} \\ \\
2 C^{\bar{i}\bar{i}\bar{i}}_{\alpha\bar{\alpha}0}
\ea
\ba{c}
2 C^{{i}{i}i}_{\alpha\bar{\alpha} 0}\\ \\
-C^{\bar{i}\bar{i}\bar{i}}_{000}
\ea
\ea
\right)
\left(
\ba{c}
T^{\bar{I}i}_{\bar{\alpha}} \\ \\T^{Ii}_{\bar{\alpha}}
\ea
\right),
\ee
\end{minipage}\\[.2cm]	

\noindent
5) If $J = \bar{J}, K = \bar{K}$, the superpotential is
\begin{equation}
W = m_I I_i \bar{I}_{\bar{i}} + \frac{1}{2} m_J (J_j)^2 + \frac{1}{2} m_K (K_k)^2
+ \lambda^{IJK} I_i J_j K_k + \lambda^{\bar{I}JK} \bar{I}_{\bar{i}} J_j K_k,
\end{equation}
and the mass matrix is
\\ \\
\begin{minipage}{16cm}
	{\bf c:}
	$\bar{I}_{\bar{i}_{\bar{\alpha}}}$,
	$J_{j_{\bar{\alpha}}}$,
	$K_{k_{\bar{\alpha}}}$,
	$I_{i_{\bar{\alpha}}}$\\[.15cm]
	{\bf r:}
	$I_{i_\alpha}$,
	$J_{j_\alpha}$,
	$K_{k_\alpha}$,
	$\bar{I}_{\bar{i}_\alpha}$\\[.2cm]
	\begin{equation}
	\left(
	\begin{array}{cccc}
	\begin{array}{c}
	m_I \\ \\
	\lambda^{\bar{I}JK} K_{k_0} C^{\bar{i}jk}_{\bar{\alpha} \alpha 0} \\ \\
	\lambda^{\bar{I}JK} J_{j_0} C^{\bar{i}jk}_{\bar{\alpha} 0 \alpha} \\ \\
	0
	\end{array}
	\begin{array}{c}
	\lambda^{IJK} K_{k_0} C^{ijk}_{\alpha \bar{\alpha} 0} \\ \\
	m_J \\ \\
	\lambda^{\bar{I}JK} \bar{I}_{\bar{i}_0} C^{\bar{i}jk}_{0 \bar{\alpha} \alpha}
	+ \lambda^{IJK} I_{i_0} C^{ijk}_{0 \bar{\alpha} \alpha} \\ \\
	\lambda^{\bar{I}JK} K_{k_0} C^{\bar{i}jk}_{\alpha \bar{\alpha} 0}
	\end{array}
	\begin{array}{c}
	\lambda^{IJK} J_{j_0} C^{ijk}_{\alpha 0 \bar{\alpha}} \\ \\
	\lambda^{\bar{I}JK} \bar{I}_{\bar{i}_0} C^{\bar{i}jk}_{0 \alpha \bar{\alpha}}
	+ \lambda^{IJK} I_{i_0} C^{ijk}_{0 \alpha \bar{\alpha}} \\ \\
	m_K \\ \\
	\lambda^{\bar{I}JK} J_{j_0} C^{\bar{i}jk}_{\alpha 0 \bar{\alpha}}
	\end{array}
	\begin{array}{c}
	0 \\ \\
	\lambda^{IJK} K_{k_0} C^{ijk}_{\bar{\alpha} \alpha 0} \\ \\
	\lambda^{IJK} J_{j_0} C^{ijk}_{\bar{\alpha} 0 \alpha} \\ \\
	m_I
	\end{array}
	\end{array}
	\right).
	\end{equation}
\end{minipage}\\[.2cm]
The F-flatness conditions are
\begin{eqnarray}
F^I_{i_0} &=& m_I \bar{I}_{\bar{i}_0} + \lambda^{IJK} I_{i_0} K_{k_0} C^{ijk}_{000}, \nonumber\\
F^{\bar{I}}_{\bar{i}_0} &=& m_I I_{i_0} +\lambda^{\bar{I}JK} \bar{I}_{\bar{i}_0} K_{k_0} C^{\bar{i}jk}_{000}.\nonumber
\end{eqnarray}		
Hence,
	\begin{equation}
	0 = \left(
	\begin{array}{cccc}
	\begin{array}{c}
	- C^{ijk}_{000} \\ \\
	X C^{\bar{i}jk}_{\bar{\alpha} \alpha 0} \\ \\
	X C^{\bar{i}jk}_{\bar{\alpha} 0 \alpha} \\ \\
	0
	\end{array}
	\begin{array}{c}
	C^{ijk}_{\alpha \bar{\alpha} 0} \\ \\
	-2 C^{ijk}_{000} \\ \\
	X C^{\bar{i}jk}_{0 \bar{\alpha} \alpha}
	+ C^{ijk}_{0 \bar{\alpha} \alpha} \\ \\
	C^{\bar{i}jk}_{\alpha \bar{\alpha} 0}
	\end{array}
	\begin{array}{c}
	C^{ijk}_{\alpha 0 \bar{\alpha}} \\ \\
	X C^{\bar{i}jk}_{0 \alpha \bar{\alpha}}
	+ C^{ijk}_{0 \alpha \bar{\alpha}} \\ \\
	-2 C^{ijk}_{0 0 0} \\ \\
	C^{\bar{i}jk}_{\alpha 0 \bar{\alpha}}
	\end{array}
	\begin{array}{c}
	0 \\ \\
	C^{ijk}_{\bar{\alpha} \alpha 0} \\ \\
	C^{ijk}_{\bar{\alpha} 0 \alpha} \\ \\
	- C^{\bar{i}jk}_{000}
	\end{array}
	\end{array}
	\right)
	\left(
	\begin{array}{c}
	T^{\bar{I}}_{\bar{i}_{\bar{\alpha}}} \\ \\
	T^{J}_{j_{\bar{\alpha}}} \\ \\
	T^{K}_{k_{\bar{\alpha}}} \\ \\
	T^I_{i_{\bar{\alpha}}}
	\end{array}
	\right).
	\end{equation}

\noindent
6) If $J = K = \bar{J} = \bar{K}$, the superpotential is
\begin{equation}
W = m_I I_i \bar{I}_{\bar{i}} + \frac{1}{2} m_K (K_k)^2
+ \lambda^{IKK} I_i K_k K_k + \lambda^{\bar{I}KK} \bar{I}_{\bar{i}} K_k K_k.
\end{equation}
The mass matrix is
	{\bf c:}
	$\bar{I}_{\bar{i}_{\bar{\alpha}}}$,
	$K_{k_{\bar{\alpha}}}$,
	$I_{i_{\bar{\alpha}}}$\\[.15cm]
	{\bf r:}
	$I_{i_\alpha}$,
	$K_{k_\alpha}$,
	$\bar{I}_{\bar{i}_\alpha}$\\[.2cm]
	\begin{equation}
	\left(
	\begin{array}{ccc}
	\begin{array}{c}
	m_I \\ \\
	\lambda^{\bar{I}KK}  K_{k_0}
	\left(C^{\bar{i}kk}_{\bar{\alpha} 0 \alpha}
	+ C^{\bar{i}kk}_{\bar{\alpha} \alpha 0}\right)  \\ \\
	0
	\end{array}
	\begin{array}{c}
	\lambda^{IKK} K_{k_0}
	\left(C^{ikk}_{\alpha 0 \bar{\alpha}} + C^{ikk}_{\alpha \bar{\alpha} 0} \right) \\ \\
	m_K \\ \\
	\lambda^{\bar{I}KK}  K_{k_0}
	\left(C^{\bar{i}kk}_{\alpha 0 \bar{\alpha}}
	+ C^{\bar{i}kk}_{\alpha \bar{\alpha} 0}\right)
	\end{array}
	\begin{array}{c}
	0 \\ \\
	\lambda^{IKK} K_{k_0}
	\left(C^{ikk}_{\bar{\alpha} 0 \alpha} + C^{ikk}_{\bar{\alpha} \alpha 0} \right)  \\ \\
	m_I
	\end{array}
	\end{array}
	\right).\nonumber
	\end{equation}
The F-flatness conditions are
\begin{eqnarray}
F^I_{i_0} &=& m_I \bar{I}_{\bar{i}_0} + \lambda^{IKK} K_{k_0} K_{k_0}
C^{ikk}_{000}, \nonumber\\
F^{\bar{I}}_{\bar{i}_0} &=& m_I I_{i_0} +\lambda^{\bar{I}KK} K_{k_0} K_{k_0} C^{\bar{i}kk}_{000},
\end{eqnarray}			
so
\\
	\begin{equation}
	0 = \left(
	\begin{array}{ccc}
	\begin{array}{c}
	- C^{ikk}_{000} \\ \\
	XC^{\bar{i}kk}_{\bar{\alpha} 0 \alpha} \\ \\
	0
	\end{array}
	\begin{array}{c}
	C^{ikk}_{\alpha 0 \bar{\alpha}}  \\ \\
	-2 C^{ikk}_{000} \\ \\
	C^{\bar{i}kk}_{\alpha 0 \bar{\alpha}}
	\end{array}
	\begin{array}{c}
	0 \\ \\
	C^{ikk}_{\bar{\alpha} 0 \alpha}  \\ \\
	- C^{\bar{i}kk}_{000}
	\end{array}
	\end{array}
	\right)
	\left(
	\begin{array}{c}
	T^{\bar{I}}_{\bar{i}_{\bar{\alpha}}} \\ \\
	T^{K}_{k_{\bar{\alpha}}} \\ \\
	T^I_{i_{\bar{\alpha}}}
	\end{array}
	\right).
	\end{equation}
	
\noindent
7) When a field does not have an $\bar{\alpha}$ or ${\alpha}$ component,
we can simply eliminate the corresponding elements in the matrix equation.
Supposing that $K$ does not contain $\bar{\alpha}$ and thus $\bar{K}$ does not contain
${\alpha}$, then in (\ref{compeqn}), the upper-left
$5\times 5$ elements in the square matrix and the upper $5$ fields in the vector are left.
If both $J,K$ have no $\bar{\alpha}$ component,
the square matrix is reduced to be $4\times 4$ and vector has only the upper $4$ fields.
If both $J$ and $K$ have no $\bar{\alpha}$ component while $I$ does not contain ${\alpha}$,
the equation is
\begin{equation}
	0 = \left(
	\begin{array}{ccc}
	\begin{array}{c}
	- C^{ijk}_{000} \\ \\ 0\\ \\
	C^{\bar{i}\bar{j}\bar{k}}_{\alpha\bar{\alpha} 0 }
	\end{array}
	\begin{array}{c}
0\\ \\
	- C^{ijk}_{000}   \\ \\
	C^{\bar{i}\bar{j}\bar{k}}_{\alpha 0\bar{\alpha} }
	\end{array}
	\begin{array}{c}
	C^{ijk}_{\bar{\alpha}\alpha 0} \\ \\
	C^{ijk}_{\bar{\alpha} 0\alpha} \\ \\
	- C^{\bar{i}\bar{j}\bar{k}}_{000}
	\end{array}
	\end{array}
	\right)
	\left(
	\begin{array}{c}
	T^{\bar{J}}_{\bar{j}_{\bar{\alpha}}} \\ \\
	T^{K}_{k_{\bar{\alpha}}} \\ \\
	T^I_{i_{\bar{\alpha}}}
	\end{array}
	\right).
	\end{equation}
	
Many of these results in this and the previous sections
have been used to solve the Goldstone modes in SUSY SO(10) models. The corresponding results will be present
in a separate publication since details of these models are not commonly known.
In the following section, we will take the SU(2) cases as examples for illustrations of our findings.

\section{CGC Identities in SU(2) Models}

The CGC identities given in the previous sections may be applied and tested directly in the SU(2) cases.
However, the CGCs we have used previously in this work differ
from what they are defined in the  angular momentum theory\cite{angular1,georgi}.
Under the traditional notations, a CGC depends strongly on the order of the
three angular momenta not only in the signs but also in the modula. In field theory
this dependence is absent.
Supposing in the SUSY models with SSB of SU(2) above the SUSY breaking scale,
if the superpotontial of the form (\ref{real1}) is taken,
it is written in the traditional notations as
\begin{equation}
W = \frac{1}{2} \sum_{I,i} M_I (I_{i_a})^2 C^{ii}_{a\bar{a};00}
+ \sum_{IJK,ijk} {\Lambda}^{IJK} I_{i_a} J_{J_b} K_{k_c} C^{ij}_{ab;kc}C^{kk}_{\bar{c}c;00},\label{realsu2}
\end{equation}
where $C^{ij}_{ab;kc}\equiv \langle k,c | i, a;j,b\rangle$ and $\bar{a}\equiv -a$.
The parameters $M$ and $\Lambda$ are proportional to $m$ and $\lambda$ of (1), respectively.

Translating the previous results into the traditional CGCs\cite{angular1,georgi}, for the real representations,
the 0-components can have a VEV while the $\pm 1$-components contribute to the Goldstone mode,
we have
\be
\left(
\ba{ccc}
C^{ij}_{0,0;k,0}\\ \\
C^{ij}_{-1,1;k,0}\\ \\
-C^{ij}_{-1,0;k,-1}
\ea
\ba{ccc}
C^{ij}_{1,-1;k,0}\\ \\
C^{ij}_{0,0;k,0}\\ \\
-C^{ij}_{0,-1;k,-1}
\ea
\ba{ccc}
-C^{ij}_{1,0;k,1}\\ \\
-C^{ij}_{0,1;k,1}\\ \\
C^{ij}_{0,0;k,0}
\ea
\right)
\left(
\ba{c}
T^I_{i_{\bar{1}}}\\ \\T^J_{j_{\bar{1}}}\\ \\T^K_{k_{\bar{1}}}
\ea
\right)=0,\label{su2real}
\ee
so that the determinant
\be
\left|
\ba{ccc}
C^{ij}_{0,0;k,0}\\ \\
C^{ij}_{-1,1;k,0}\\ \\
-C^{ij}_{-1,0;k,-1}
\ea
\ba{ccc}
C^{ij}_{1,-1;k,0}\\ \\
C^{ij}_{0,0;k,0}\\ \\
-C^{ij}_{0,-1;k,-1}
\ea
\ba{ccc}
-C^{ij}_{1,0;k,1}\\ \\
-C^{ij}_{0,1;k,1}\\ \\
C^{ij}_{0,0;k,0}
\ea
\right|=0.\label{su2real2}
\ee
For special cases, the corresponding matrix becomes
\be
\left(
\ba{cc}
C^{ij}_{0,0;j,0} &~~[1+(-1)^{i+2j}] C^{ij}_{1,-1;j,0}\\
C^{ij}_{-1,1;j,0} &~~C^{ij}_{0,0;j,0}-C^{ij}_{0,1;j,1}
\ea
\right)\,,
\ee
when $J=K$ in (\ref{su2real}),
and
\be
\left(
\ba{cc}
C^{ii}_{0,0;i,0} &~~[1+(-1)^{3i}] C^{ii}_{1,-1;i,0}\\
C^{ii}_{-1,1;i,0} &~~C^{ii}_{0,0;i,0}-C^{ii}_{0,1;i,1}
\ea
\right)\,,
\ee
when $I=J=K$, following the identities\cite{angular1}
\be
C^{i_1j_1}_{m_1,m_2;j,m}=(-1)^{j_1+j_2-j}C^{j_1i_1}_{m_2,m_1;j,m}, ~~
C^{i_1j_1}_{m_1,m_2;j,m}=(-1)^{j_2+m_2}\sqrt{\frac{2j+1}{2j_2+1}}C^{i_1j}_{m_1,-m;j_2,-m_2}.
\ee.
The determinants of all these matrices are zero.

In SU(2), although  a representation of dimension $2n+1$
is self-conjugate for an integer $n$, it need to be taken as complex when it carries an extra charge.
Examples of this kind include the SSB of $SU(2)_R\times U(1)_{B-L}$ into $U(1)_Y$ and
of $SU(2)_L\times U(1)_{Y}$ into $U(1)_{EM}$,
although the latter SSB happens below the SUSY breaking scale and is not relevant to
the present study.
So we focus on the superpotential of the form (\ref{complex1}).
Supposing a field as $I$ which is dimension $2i+1$ under SU(2)
and its $i_0$ component develops a VEV,
we get the identity
\be
\left|
\ba{ccc|ccc}
\ba{c}
C^{ij}_{i_0,j_0;k,k_0}\\ \\
0\\ \\
0\\ \\
0\\ \\
C^{ij}_{i_+,j_-;k,k_0}\\ \\
-C^{ij}_{i_+,j_0;k,k_+}
\ea
\ba{c}
0\\ \\
C^{ij}_{i_0,j_0;k,k_0}\\ \\
0\\ \\
C^{ij}_{i_-,j_+;k,k_0}\\ \\
0\\ \\
-C^{ij}_{i_0,j_+;k,k_+}
\ea
\ba{c}
0\\ \\
0\\ \\
C^{ij}_{i_0,j_0;k,k_0}\\ \\
-C^{ij}_{i_-,j_0;k,k_-}\\ \\
-C^{ij}_{i_0,j_-;k,k_-}\\ \\
0
\ea
\ba{c}
0\\ \\
C^{ij}_{i_-,j_+;k,k_0}\\ \\
-C^{ij}_{i_-,j_0;k,k_-}\\ \\
C^{ij}_{i_0,j_0;k,k_0}\\ \\
0\\ \\
0
\ea
\ba{c}
C^{ij}_{i_+,j_-;k,k_0}\\ \\
0\\ \\
-C^{ij}_{i_0,j_-;k,k_-}\\ \\
0\\ \\
C^{ij}_{i_0,j_0;k,k_0}\\ \\
0
\ea
\ba{c}
-C^{ij}_{i_+,j_0;k,k_+}\\ \\
-C^{ij}_{i_0,j_+;k,k_+}\\ \\
0\\ \\
0\\ \\
0\\ \\
C^{ij}_{i_0,j_0;k,k_0}
\ea
\ea
\right|=0,\label{6x6su2}
\ee
where $i_\pm =i_0\pm 1$, {\it etc.}.
Note that in (\ref{6x6su2}), $i_0,j_0,k_0=i_0+j_0$ take all allowed values
therefore this identity is very general.

Most of the special cases follow the general discussions in the previous sections.
For $J,K$ does not contain $\bar{\alpha}$, {\it i.e.} $j_0=-j, k_0=-k$,
$I$ does not contain ${\alpha}$ following the triangle rule and $i+j=k$,
(\ref{6x6su2}) reduces to a $3\times 3$ whose determinant turns out to be
\be
1- (C^{ij}_{i,j-1;i+j,i+j-1})^2- (C^{ij}_{i-1,j;i+j,i+j-1})^2\equiv 0 \,,
\ee
which is simply true.

We believe the approach we present above is the easiest way in deriving the identities (\ref{su2real2},\ref{6x6su2}).
Numerically we have tested these identities to be true
In principle, the identities in (\ref{su2real2},\ref{6x6su2}) can have a proof
using the traditional approach in angular momentum theory,
which is, although desirable, not our focus in the present work.

\section{Summary}

In SUSY models, the components in the Goldstone modes are shown to be
explicitly proportional to the VEVs,
with no more dependence on the parameters in the superpotential,
which are consistent with the general results on gauge theories with SSB.
Identities among the CGCs are found as examinations to
the calculations and are applied to the SU(2) cases.

DXZ thank Prof. Y.-X. Liu for helpful discussions.

\end{document}